# Low-field quantum Hall transport in an electron Fabry-Perot interferometer: Determination of constriction filling vs front-gate voltage

Ping V. Lin, F. E. Camino*, and V. J. Goldman
*Department of Physics, Stony Brook University, Stony Brook, NY 11794-3800, USA*

We report systematic quantum Hall transport experiments on Fabry-Perot electron interferometers at ultra-low-temperatures. The GaAs/AlGaAs heterostructure devices consist of two constrictions defined by etch trenches in 2D electron layer, enclosing an approximately circular island. Front gates deposited in etch trenches allow to fine-tune the device for symmetry and to change the constriction filling, relative to the bulk. The low-field longitudinal and Hall magnetotransport shows Shubnikov-de Haas oscillations and integer quantum Hall plateaus. A systematic variation of front-gate voltage affects the constriction and the island electron density, while the bulk density remains unaffected. This results in quantized plateaus in longitudinal resistance, while the Hall resistance is dominated by the low-density, low-filling constriction. At lower fields, when the quantum Hall plateaus fail to develop, we observe bulk Shubnikov-de Haas oscillations in series corresponding to an integer filling of the magnetoelectric subbands in the constrictions. This indicates that the whole interferometer region is still quantum-coherent at these lower fields at 10 mK. Analyzing the data within a Fock-Darwin model, we obtain the constriction electron density as a function of the front-gate bias and, extrapolating to the zero field, the number of electric subbands (conductance channels) resulting from the electron confinement in the constrictions.

## I. INTRODUCTION

There has been a continuing wealth of research into the ground state and transport properties of confined two-dimensional (2D) electron systems ever since the discovery of the integer quantum Hall effect[1] and development of lithographic techniques. The integer quantum Hall effect (IQHE) can be understood in terms of transport by one-dimensional (1D) chiral edge channels corresponding to an integer number of fully occupied Landau levels.[2-4] In this picture, near an integral Landau level filling $\nu \approx f$, when the chemical potential lies in the gap of localized bulk states, the current is carried by dissipationless edge channels and the Hall resistance is quantized exactly to $h/fe^2$. Dissipative transport occurs when current is carried by the extended bulk states of the partially occupied topmost Landau level, between the plateaus. Such interpretation of the IQHE of non-interacting electrons in terms of edge channels is straightforward since for non-interacting electrons the chiral edge channels are formed in one-to-one correspondence with the bulk Landau levels. Including effects of electron interaction is not so straightforward, but, qualitatively, the concept of current-carrying chiral edge channels is still applicable.[5-10]

In a constricted geometry, even in zero magnetic field $B = 0$, an approximate quantization of conductance[11,12] is understood as resulting from size-quantized non-chiral 1D conducting channels passing through the constriction.[13] In a quantizing $B$, the size-quantized and the chiral edge channels hybridize, there exists a transitional regime where both effects co-exist, and the plateau positions in $B$ depend on both size and Landau quantization. Here, the non-interacting electron theory does not provide quantitatively accurate description, so that effects of interaction resulting in a self-consistent confining potential have to be included. In addition, in such constrictions, "backscattering" by quantum tunneling between the extended edge states is possible and leads to a deviation from exact plateau quantization.

In this paper we present a comprehensive experimental characterization of quantum Hall (QH) and Shubnikov-de Haas (SdH) transport in an electron Fabry-Perot interferometer.[14-16]

Similar electron interferometer devices have been studied by others in the integer QH regime.[17-19] These studies are moreover motivated by application of such interferometer devices in the fractional QH regime, where interference of fractionally-charged Laughlin quasiparticles has been reported.[20-25] Additional motivation is provided by proposed application of such Fabry-Perot interferometers, in conjunction with quantum antidots,[26] to detection of non-Abelian braiding statistics and as a physical implementation of topological quantum computation.[27-30]

## II. SAMPLES AND EXPERIMENTAL TECHNIQUES

The interferometer sample was fabricated from a very low disorder double-δ-doped GaAs/AlGaAs heterostructure.[31] The 2D electron system is buried 320 nm below the surface. First, Ohmic contacts are formed on a pre-etched mesa. Then etch trenches are defined by electron-beam lithography, using proximity correction software for better definition of narrow and long gaps between the exposed areas. After a shallow 160 nm wet etch, 50 nm thick Au/Ti front-gate metallization is deposited in a self-aligned process. Finally, samples are mounted on sapphire substrates with In metal, which serves as the global back gate. The interferometer sample studied in this paper is the same as in Ref. 24, but on a subsequent cool-down and under different illumination.

Samples were cooled in the tail of the mixing chamber of a top-loading into mixture dilution $^3$He-$^4$He refrigerator. A bulk 2D electron density $n_B = 1.16 \times 10^{11}$ cm$^{-2}$ was achieved after illumination by a red LED at 4.2 K. All experiments reported in this work were performed at the fixed bath temperature of 10 mK, calibrated by nuclear orientation thermometry. Extensive cold filtering in the electrical leads attenuates the electromagnetic background "noise" incident on a sample, allowing to achieve effective electron temperatures of ≤15 mK.[23] Four-terminal longitudinal $R_{XX} = V_X / I_X$ and Hall $R_{XY} = V_Y / I_X$ magnetoresistance, see Fig. 1, were measured with a lock-in technique at 5.4 Hz. The excitation current was set so as to keep the larger, Hall or longitudinal voltage ≤5 μV.

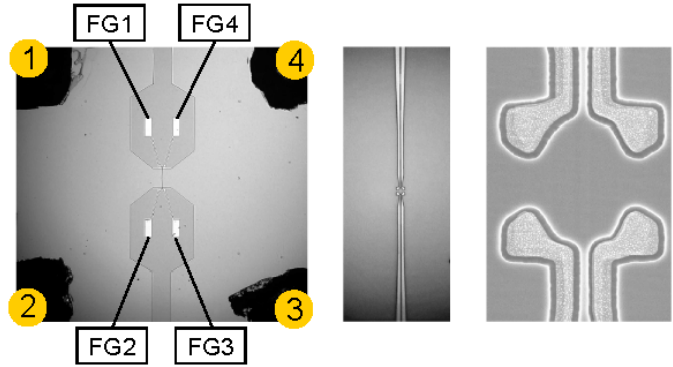

FIG. 1. A Fabry-Perot electron interferometer device. Optical (two left) and scanning electron (SEM, right) micrographs of the interferometer sample. Numbered circles on the four corners of the 4×4 mm mesa show Ohmic contacts to 2D bulk electron layer. Four front gates (FG1-4) are deposited in shallow etch trenches, defining a circular island separated from the 2D bulk by two 1.2 μm wide constrictions. In a quantizing magnetic field, chiral edge channels follow an equipotential at the periphery of the undepleted 2D electrons. Longitudinal $R_{XX}$ (current 1-4, voltage 2-3) and Hall $R_{XY}$ (current 2-4, voltage 1-3) resistances are measured. The back gate (not shown) extends over the entire sample.

## III. EXPERIMENTAL RESULTS AND ANALYSIS
### A. Magnetotransport

Figures 2 - 4 summarize experimental longitudinal and Hall four-terminal magnetoresistance in sample M97Ce, the same as reported in Ref. 24, but under different illumination, taken in a range of front-gate voltages $-580 \leq V_{FG} \leq +100$ mV. Even at zero front-gate $V_{FG} = 0$, the GaAs surface depletion of the etch trenches, which remove the doping layer, creates electron confining potential, so that the constriction and the island electron densities are less than the 2D "bulk". Application of a negative $V_{FG}$ depletes the constrictions-island region of the sample further. The two constrictions were tuned for approximate symmetry by application of a constant ±20 mV differential bias between FG1 and FG4, additional to the common front-gate bias given in this paper as $V_{FG}$. Detuning front-gate voltage from symmetry allows to verify each constriction filling separately.

Because in a uniform applied $B$ the Landau level filling factor $\nu = hn/eB$ is proportional to the local electron density, $\nu$ in the depleted regions of the sample is different from the 2D bulk $\nu_B$. While $\nu \propto n/B$ is a variable, the quantum Hall exact filling $f$ is a quantum number defined by the quantized Hall resistance as $f = h/e^2 R_{XY}$. Because QH plateaus have finite width, regions with different $\nu$ may have the same $f$. In samples with lithographic constrictions, in general, there are two possibilities: (i) when depletion is small and on a wide QH plateau, the whole sample may have the same QH filling $f$; and (ii) more often, the constriction filling $f_C$ and the bulk filling $f_B$ are different. As can be seen in Fig. 2, as the front gates are biased more negative, there is a continuous series of well-developed constriction QH plateaus for each $f_C$, shifting to lower magnetic fields, and thus to higher $f_B$ plateaus.

The Hall resistance $R_{XY}$ allows us to determine the filling in the constrictions, the plateau positions in $B$ giving definitive values of $f_C$. The longitudinal $R_{XX}$ shows quantum Hall minima and quantized plateaus at $R_{XX} = (h/e^2)(1/f_C - 1/f_B)$, when plateaus in constrictions and the bulk overlap in $B$.[32] Note the special case: when $f_C = f_B$, $R_{XX} = 0$. Thus, a quantized plateau in $R_{XX}(B)$ implies quantum Hall plateaus for both the constriction region and the bulk, and a set of quantized $R_{XX}(B)$ plateaus provides definitive values for both $f_C$ and $f_B$. Evolution of several stronger QH constriction and bulk plateaus as a function of $V_{FG}$ are indicated in Fig. 3. As expected, the constriction plateaus are shifted to lower magnetic fields by a negative front-gate voltage, while the bulk plateaus are not affected.

Figure 4 shows detail of the magnetotransport data in the range of $B$ where SdH oscillations in the bulk occur, transitioning to developing QH plateaus. As seen in the raw data, SdH oscillations are not shifted in $B$ by front-gate bias; this is confirmed by Fourier analysis, which gives a $V_{FG}$-independent SdH oscillation frequency corresponding to the bulk density. The vertical positions of the bulk SdH oscillations are grouped in series corresponding to the number of conduction channels passing through the constriction. This shows as a constant resistance plateau in the Hall data, and a $(h/e^2)(1/f_C - 1/f_B)$ background, the bulk Hall effect when $f_C$ is constant, in the $R_{XX}$ data. The values of the Hall plateaus and the $B = 0$ intercept of the negative slope $R_{XX}$ background are both $h/f_C e^2$, which can be used to ascertain the channel number of the constriction series.

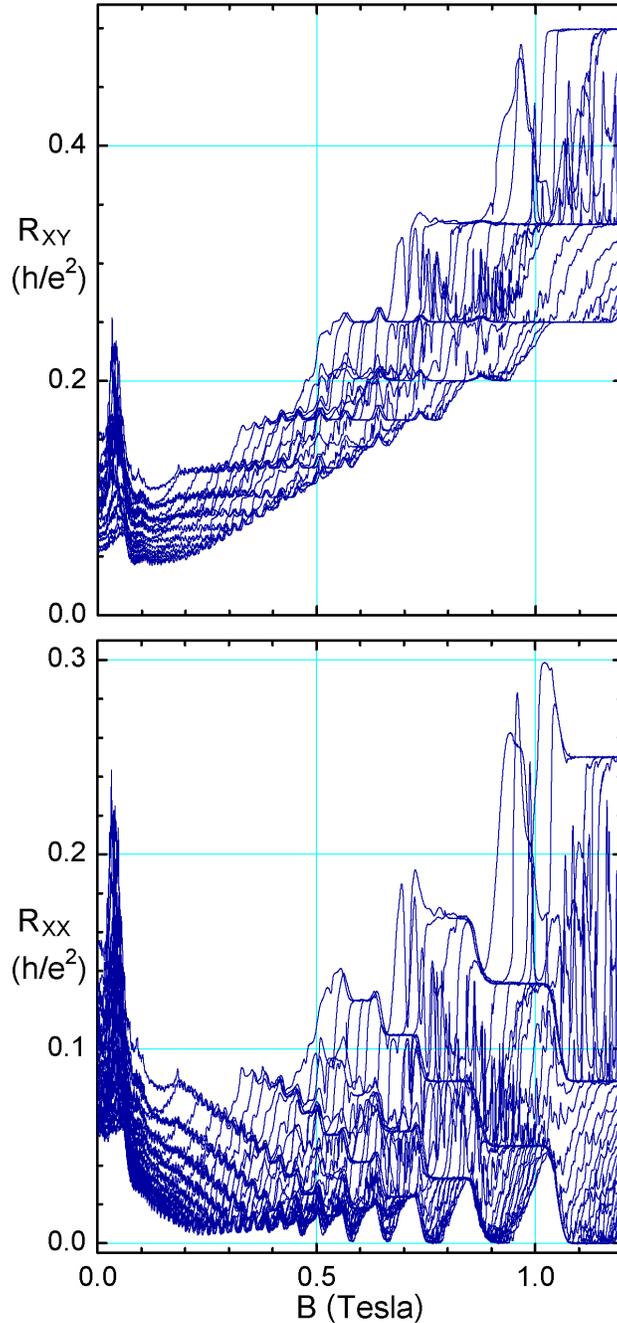

FIG. 2. Representative longitudinal and Hall magnetoresistance traces of the interferometer sample. The front-gate voltage is stepped by multiples of 20 mV in the range $-580 \leq V_{FG} \leq +100$ mV. The lowest traces correspond to the positive bias, the higher resistance (lower electron density) to the negative $V_{FG}$. The zero resistance level is the same for all traces. Application of $V_{FG}$ changes the electron density in the interferometer region, both the island and the constrictions, thus shifting the $B$-positions of the quantized plateaus. The smallest filling

factor, that in constrictions, determines the Hall signal, while the longitudinal signal depends on filling in all regions of the sample, including the 2D bulk.

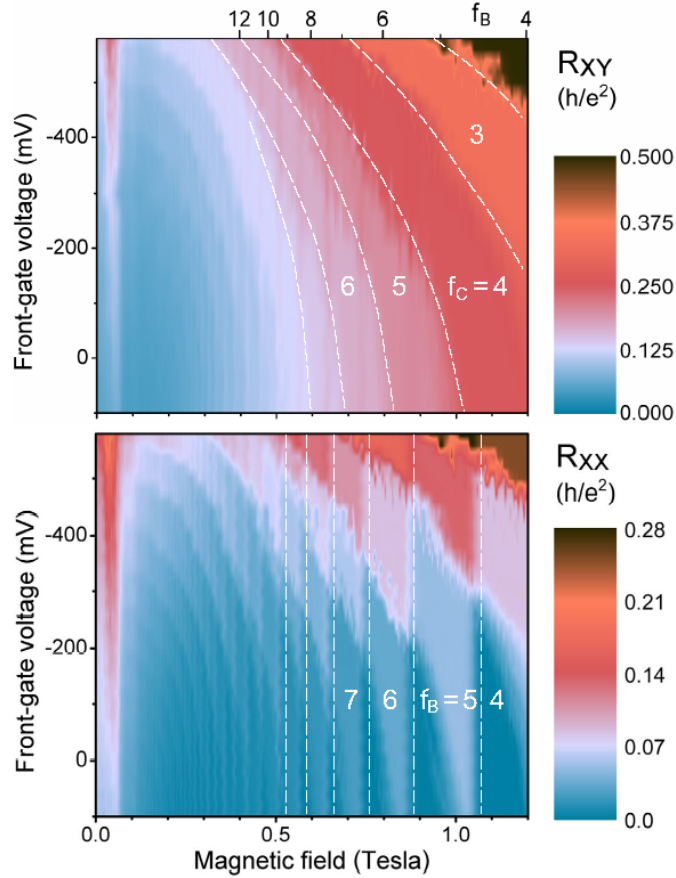

FIG. 3. Color-mapped plot of the magnetoresistance data of Fig. 2. The Hall $R_{XY}$ and longitudinal $R_{XX}$ plateau regions correspond to the same shade. Note that the constriction plateaus are shifted to lower magnetic fields by a negative front-gate voltage, while the bulk plateaus are not affected. The absolute resistance values of the $R_{XX}$ plateaus allow to determine both constriction and bulk fillings as a function of magnetic field, as shown. The dashed white lines give approximate boundaries between consecutive QH plateaus, and are guides for the eye.

It is not surprising that the QH edge channels pass through both constrictions. As can be seen in Fig. 4, there is a smooth, continuous transition from well-developed QH constriction plateaus to the low-field magnetoelectric conduction channel regime for each $f_C$ series formed by various $V_{FG}$ traces. This means that the whole interferometer region, including both constrictions, is quantum-coherent even at 0.1 T, and most likely, indeed, even at $B = 0$. If the two constrictions were not quantum-coherent, their individual resistances would add, which would be seen as an apparent doubling of the constriction channel number as $B$ is lowered.

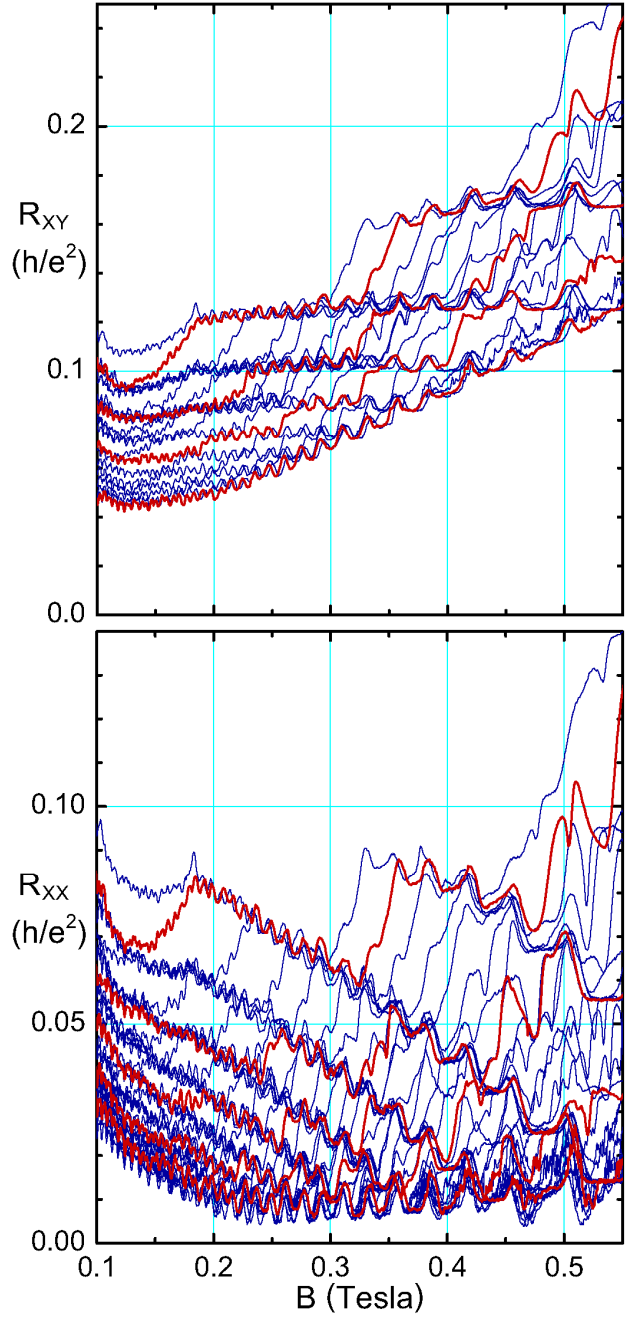

FIG. 4. Blow-up of the magnetoresistance data of Fig. 2 in the region of Shubnikov-de Haas oscillations and developing quantum Hall plateaus in the bulk. Some traces are shown in thicker red lines to help distinguish individual traces. The lowest traces correspond to the positive bias, the higher resistance (lower electron density) - to the negative $V_{FG}$. Note that the $B$-positions of the bulk SdH oscillations are not affected by $V_{FG}$, while superimposed on resistance background determined by the number of the conduction channels in the constrictions, which is shifted by $V_{FG}$. This allows to separate the bulk and constriction features. The zero resistance level is the same for all traces.

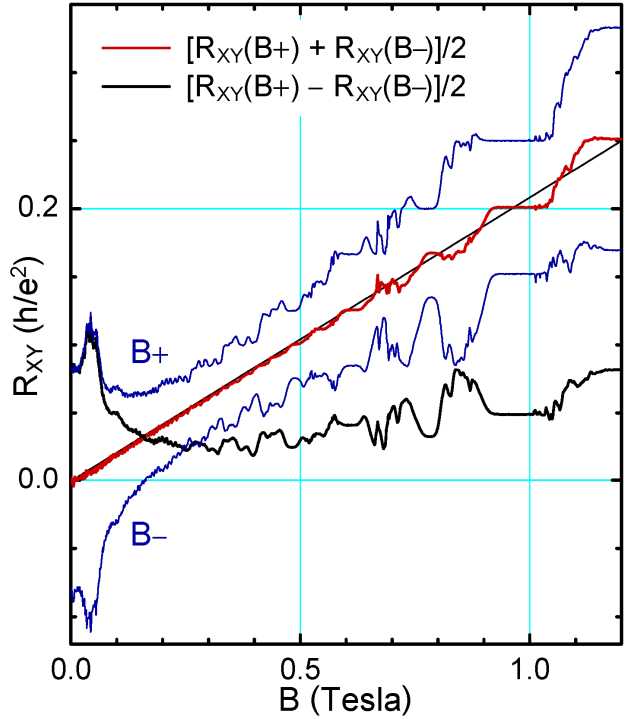

FIG. 5. Experimental four-terminal Hall $R_{XY}$ also contain longitudinal contribution. The two directly measured traces shown (thin blue lines) are obtained with magnetic field up ($B+$) and down ($B-$). The $R_{XY}(B-)$ trace is shown multiplied by $-1$, both horizontal and vertical axes. The middle trace (thick red line) is the average $\frac{1}{2}[R_{XY}(B+)+R_{XY}(B-)]$, which, according to Onsager relations, gives the true bulk $R_{XY}$ (straight thin line gives the bulk density). Likewise, the difference $\frac{1}{2}[R_{XY}(B+)-R_{XY}(B-)]$ gives the longitudinal $R_{XX}$, which displays the quantized plateaus, e.g., $R_{XX} = 0.05\, h/e^2$ ($f_C = 4$, $f_B = 5$) at $B \approx 0.98$ T. Data taken at $V_{FG} = -260$ mV.

Because a four-terminal $R_{XY}$ generally contains longitudinal contributions, it may not be clear-cut as to what is the true Hall effect. We can ascertain the assignment of various features to the bulk or to the constriction by the following two techniques. First, we can reverse the direction of the magnetic field, that is, take the corresponding magnetoresistance data at both $B+$, up, and $B-$, down, shown in Fig. 5. The $R_{XY}(B-)$ data is multiplied by $-1$, both the magnetic field and resistance. According to Onsager relations for a magnetoconductivity tensor in an inversion-symmetric sample,[33] the Hall contribution changes sign, while the diagonal contributions remain unaffected. Thus, the average Hall $\frac{1}{2}[R_{XY}(B+)+R_{XY}(B-)]$ corresponds to the true bulk Hall effect, with all longitudinal contributions to resistance, including the effect of constrictions, removed (within the experimental accuracy). The thin solid line in Fig. 5 gives the classical Hall slope corresponding to the bulk density $n_B = 1.16 \times 10^{11}$ cm$^{-2}$, obtained from the $B$ positions of the $V_{FG}$-independent QH plateaus in Figs. 2 and 3; as can be seen, it matches the $B+$, $B-$ average slope well. The difference $\frac{1}{2}[R_{XY}(B+)-R_{XY}(B-)]$ has no Hall contribution, and closely follows the raw $R_{XX}$ data at the same front-gate voltage. Such analysis for the low-filling fractional QH regime has been reported in Ref. 34.

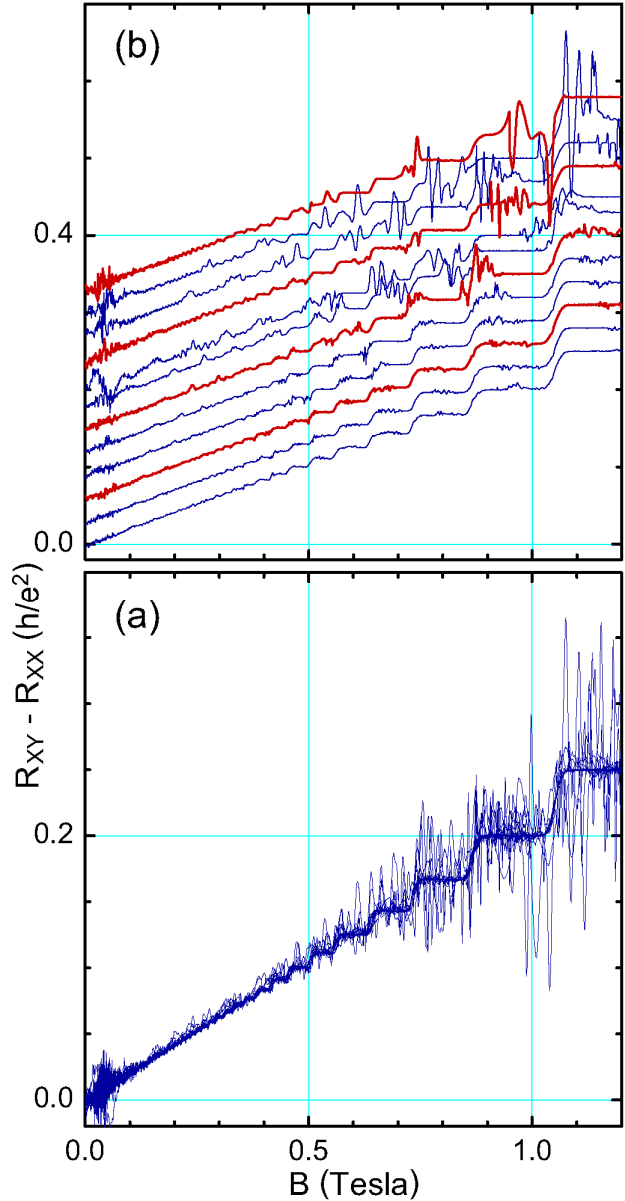

FIG. 6. Representative traces illustrating subtraction of experimental Hall and longitudinal magnetoresistance at the same front-gate voltage. The lower panel shows the difference traces at various $V_{FG}$, all having true zero level. The subtraction results in the bulk Hall resistance (darker central region), with superimposed features due to mesoscopic effects and tunneling in the constrictions, different in each individual $V_{FG}$ trace. The upper panel shows several individual $V_{FG}$ traces shifted vertically by $0.01\,h/e^2$ per $-20$ mV of $V_{FG}$.

The second technique is approximate; it is exact in certain bulk-edge network models of QH transport,[35-38] and is also an approximate semiclassical result in the limit of $\sigma_{XX} \ll \sigma_{XY}$ in 2D samples.[39] For each $V_{FG}$ in Fig. 2, we subtract longitudinal from the Hall magnetoresistance, $R_{XY}(B) - R_{XX}(B)$, both for $B+$. When both constriction and bulk are on a QH plateau, it is apparent that the difference is $h/f_B e^2$, the bulk Hall effect. However, this technique also subtracts the finite $\sigma_{XX}$ contributions between the plateaus, present both in $R_{XY}$ and $R_{XX}$, as shown in Fig. 6.

The fine structure in the traces of Figs. 2 - 6 is attributed to disorder-assisted tunneling and quantum interference effects. It is particularly visible in the difference data of Fig. 6 (more constricted sample, larger magnitude negative $V_{FG}$), since the individual $R_{XY}$ and $R_{XX}$ traces for the same $V_{FG}$ were taken several days apart, so that the detailed $B$-positions and magnitude of the "mesoscopic features" do not match, and thus do not subtract, due to their slow drift as a function of time. Aharonov-Bohm oscillations,[13-18] present in some data, have small amplitude ($\leq 4 \times 10^{-3}$ $h/e^2$) and are not visible on the scale of Figs. 2 - 6.

### B. Constriction electron density

The $B=0$, $V_{FG}=0$ shape of the electron density profile resulting from etch trench depletion in the interferometer region of the sample is illustrated in Fig. 1 of Ref. 24. The interferometer island is large, contains 2 - 4×10³ electrons, and the 2D electron density profile is determined mostly by the classical electrostatics, minimizing the energy of electron-electron repulsion, compensated by attraction to the positively charged donors. The Fabry-Perot device depletion potential has saddle points in the constrictions, and so has the resulting electron density profile. In a quantizing magnetic field edge channels form, but the overall electron density profile closely follows the $B=0$ profile in these relatively large devices, so as to minimize total Coulomb energy.

Because the in-plane screening by 2D electrons is relatively weak,[6,7,10] application of a negative front-gate voltage $V_{FG}$ decreases electron density throughout the interferometer region. The main depletion is provided by the etch trenches; modeling[14,16,22] shows that application of a moderate $V_{FG}$, besides the overall depletion, increases effective depletion length by ~100 nm/V. Changing magnetic field affects the equilibrium electron density profile in the device only weakly, particularly for $f \geq 4$, the principal effect is to redistribute the electron occupation between various Landau levels. In a fixed $B$, when the density of states in each Landau level in a given area is fixed also, application of $V_{FG}$ changes occupation of these states.

We model the constriction following non-interacting electrons Fock-Darwin model,[40-42] as a 1D conductor with a parabolic confining potential with energy level (1D subbands) spacing $\hbar\omega_0$. In a quantizing magnetic field with cyclotron energy $\hbar\omega_C$, hybrid magnetoelectric subbands with bottom at energy $E_n = (n+\frac{1}{2})\sqrt{(\hbar\omega_0)^2 + (\hbar\omega_C)^2}$, where $n = 0, 1, 2, \ldots$, serve as conduction channels. In GaAs, spin splitting of the subbands is small compared with $\hbar\omega_C$, and develops only at higher magnetic fields. As $\hbar\omega_C \propto B$ is increased from zero, these magnetoelectric subbands cross the chemical potential, become de-populated, and so the number of the constriction conduction channels decreases. We use the experimental number of constriction conduction channels, taken as $f_C$ from the corresponding constriction plateau position (the exact filling $\nu = f$) in the magnetoresistance data in Fig. 2, shown as circles in Fig.

7(a). Both $R_{XX}(B)$ and $R_{XY}(B)$ data sets yield consistent constriction plateau positions; the horizontal error bars represent the uncertainty in the $B$-position of the centers of the plateaus.

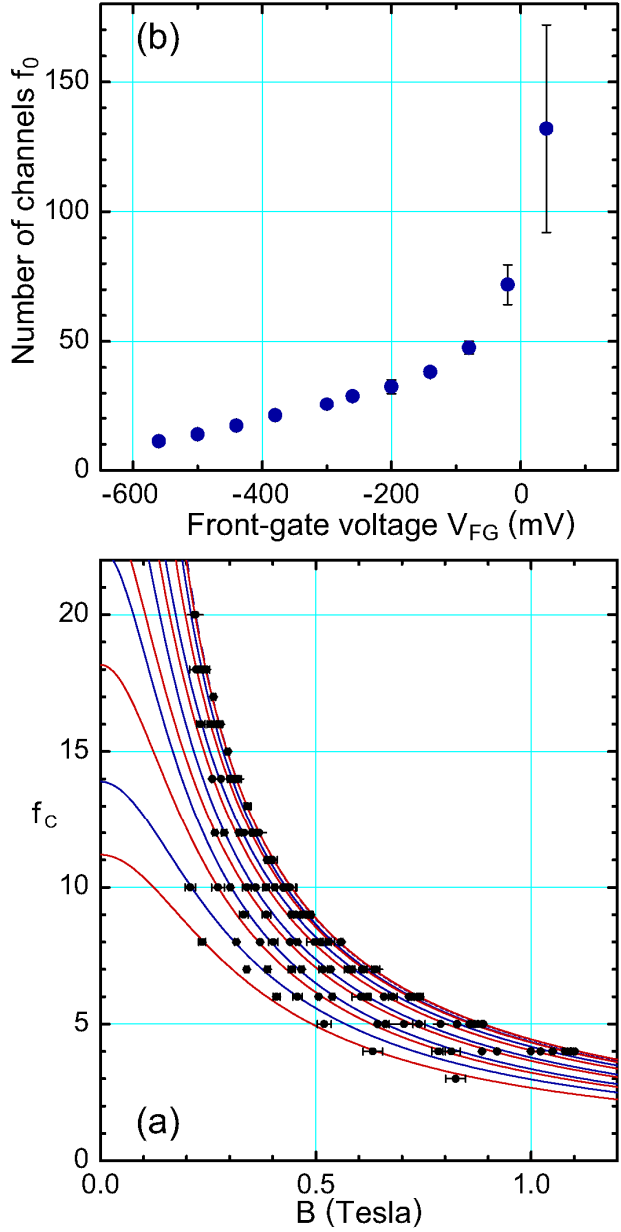

FIG. 7. (a) Positions of the constriction plateaus from the data of Fig. 2 (circles with horizontal error bars) and the fits to the 1D Fock-Darwin model (lines). Each set of points and the fit correspond to a particular $V_{FG}$. (b) Zero-field number of conduction channels in the constriction obtained from the fits shown in (a).

Each set of channel number $\{f, B_f\}$ points, corresponding to a particular $V_{FG}$, is fitted with

$$f = f_0 / \sqrt{1 + (f_0 B_f / B_1)^2}, \tag{1}$$

where $f_0$ is the conduction channel number at $B=0$, and $B_1$ is the $f=1$ plateau center, both refer to constriction. The fits, shown in Fig. 7(a), were performed for the data for $V_{FG}$ stepped typically by 60 mV; they yield $f_0$ as the best fit parameter, plotted in Fig. 7(b). One $f_0 = 582 \pm 300$ for $V_{FG} = +100$ mV is not shown in Fig. 7(b). The value of $f_0 \sim 100$ at zero front-gate bias is consistent with the four-terminal sample resistance of ~300 Ohm at 1.2 K. (At mK temperatures, the $B=0$ sample resistance is higher, ~1.2 kOhm, as can be seen in Fig. 2, likely due to quantum interference effects in the device). The absolute error in $f_0$ is, unsurprisingly, large for nearly open constriction, when the constriction density is only slightly less than $n_B$.

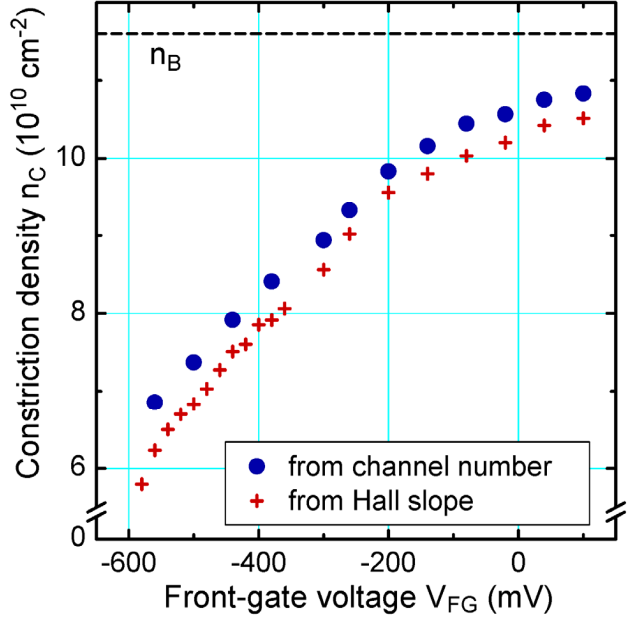

FIG. 8. Constriction electron density obtained from the conduction channel number analysis, Eq. (1), in Fig. 7(a) (circles). Also shown is $n_C$ obtained by conventional Hall slope (forced through zero) analysis (crosses). This neglects confinement in the constriction, and thus systematically underestimates the density. The 2D bulk density $n_B$ is shown by the dashed line.

The second fit parameter, $B_1$, gives information on constriction density, $n_C = eB_1/h$ at $\nu_C = 1$. Thus obtained constriction density is plotted as a function of $V_{FG}$ in Fig. 8. Since $n_C$ is derived from the QH transport data, it should correspond to the electron density near the saddle point in the constriction, which determines the constriction QH filling. We also have determined the constriction density as $n_C = \nu eB/h$, from the raw transport data Hall slope (crosses in Fig. 8). The classical Hall line is forced through zero at $B=0$, which results in a systematically larger Hall slope, and so underestimates $n_C$. Roughly, this procedure is equivalent to the Fock-Darwin analysis described above, but setting $f_0 \to \infty$ in Eq. (1).

Within Fock-Darwin model, that is, assuming parabolic constriction confinement potential and neglecting electron interaction, we can also estimate the constriction width[43,44] at chemical potential as $W \approx f_0/2\sqrt{2n_C}$, where one factor of 2 account for electron spin. The resulting width varies in the range $160 \leq W \leq 6,000$ nm in the experimental range of $V_{FG}$. The larger value, 6 µm, is much larger than the lithographic constriction with of 1.2 µm. Both assumptions of the model are not realistic, and it is remarkable that some values obtained, such as $f_0$ and $n_C$ are reasonable, while others, such as $W$, are not.

Several words are in order regarding the island center electron density $n_I$. Etch trench depletion modeling at $V_{FG} = 0$ gives $n_I$ ~2% lower than $n_B$, and ~7% greater than constriction saddle point density $n_C$.[24,45] This is consistent with the results shown in Fig. 8, giving $n_C \approx 0.92 n_B$ at $V_{FG} = 0$. While present QH transport experiments do not probe $n_I$, an analysis of $V_{FG}$-dependence of the period of the Aharonov-Bohm oscillations at lower filling $f \leq 4$ integer QH plateaus has lead us to conclude that $n_C$ decreases proportionately less than $n_I$, upon application of a negative $V_{FG}$.[16,22] This experimental conclusion is counterintuitive, but can be understood if one considers that the front gates have long leads and surround the island, while being only to one side of a constriction, Fig. 1. Accordingly, the island QH edge channels, which follow the constant electron density contours with density equal that in the constrictions, move inward, towards the island center, the interference path area shrinks, and the Aharonov-Bohm period increases.

## IV. CONCLUSIONS

In conclusion, we have presented a comprehensive experimental characterization of quantum Hall and Shubnikov-de Haas transport in an electron Fabry-Perot interferometer. We find that application of front-gate voltage affects the constriction electron density, while the bulk density remains unaffected. This results in quantized plateaus in longitudinal resistance, while the Hall resistance is dominated by the low-density, low-filling constriction. At lower fields, when the quantum Hall plateaus fail to develop, we observe bulk Shubnikov-de Haas oscillations in series corresponding to an integer filling of the magnetoelectric subbands in the constriction. From a Fock-Darwin analysis, we obtain the constriction electron density as a function of the front-gate bias and, the zero-field number of 1D electric subbands (conductance channels), resulting from the electron confinement in the constrictions.

## ACKNOWLEDGMENTS

We thank Dr. Wei Zhou for help in experiments. This work is supported in part by the National Science Foundation under grant DMR-0555238.